\documentclass[prl,twocolumn,floatfix]{revtex4}
\usepackage[dvips]{graphicx}
\usepackage{latexsym}
\usepackage{amsmath}
\usepackage{amsfonts}
\usepackage{amssymb}
\usepackage{bm}
\usepackage{color}
\begin{document}
	\newcommand{\fig}[2]{\includegraphics[width=#1]{#2}}
	\newcommand{\la}{{\langle}}
	\newcommand{\ra}{{\rangle}}
	\newcommand{\dg}{{\dagger}}
	\newcommand{\upa}{{\uparrow}}
	\newcommand{\dna}{{\downarrow}}
	\newcommand{\ab}{{\alpha\beta}}
	\newcommand{\ias}{{i\alpha\sigma}}
	\newcommand{\ibs}{{i\beta\sigma}}
	\newcommand{\hH}{\hat{H}}
	\newcommand{\hn}{\hat{n}}
	\newcommand{\hc}{{\hat{\chi}}}
	\newcommand{\hU}{{\hat{U}}}
	\newcommand{\hV}{{\hat{V}}}
	\newcommand{\br}{{\bf r}}
	\newcommand{\bk}{{{\bf k}}}
	\newcommand{\bq}{{{\bf q}}}
	\def\gsim{~\rlap{$>$}{\lower 1.0ex\hbox{$\sim$}}}
	\setlength{\unitlength}{1mm}
	\newcommand{{\vhf}}{\chi^\text{v}_f}
	\newcommand{{\vhd}}{\chi^\text{v}_d}
	\newcommand{{\vpd}}{\Delta^\text{v}_d}
	\newcommand{{\ved}}{\epsilon^\text{v}_d}
	\newcommand{{\vved}}{\varepsilon^\text{v}_d}
	\newcommand{{\tr}}{{\rm tr}}
	\newcommand{\pprl}{Phys. Rev. Lett. \ }
	\newcommand{\pprb}{Phys. Rev. {B}}

\title{Quantum anomalous vortex and Majorana zero mode in iron-based superconductor Fe(Te,Se)}
\author{Kun Jiang,$^{1,2}$, Xi Dai,$^{3}$ and Ziqiang Wang$^{1}$}
\affiliation{$^1$ Department of Physics, Boston College, Chestnut Hill, MA 02467, USA}
\affiliation{$^2$ Beijing National Laboratory for Condensed Matter Physics and Institute of Physics,
	Chinese Academy of Sciences, Beijing 100190, China}
\affiliation{$^3$ Department of Physics, Hong Kong University of Science and Technology, Kowloon, Hong Kong}
\date{\today}

\begin{abstract}
In topological insulators doped with magnetic ions, spin-orbit coupling and ferromagnetism give rise to the quantum anomalous Hall effect. Here we show that in s-wave superconductors with strong spin-orbit coupling, magnetic impurity ions can generate topological vortices in the absence of external magnetic fields. Such vortices, dubbed quantum anomalous vortices, support robust Majorana zero-energy modes when superconductivity is induced in the topological surface states. We demonstrate that the zero-energy bound states observed in Fe(Te,Se) superconductors are possible realizations of the Majorana zero modes in quantum anomalous vortices produced by the interstitial magnetic Fe. The quantum anomalous vortex matter not only advances fundamental understandings of topological defect excitations of Cooper pairing, but also provides new and advantageous platforms for manipulating Majorana zero modes in quantum computing.
\end{abstract}
\maketitle

\textit{Introduction} Harvesting localized Majorana fermion excitations has thrived in condensed matter and materials physics for both its fundamental value and its potential for fault-tolerant nonabelian quantum computing \cite{read,kitaev,ivanov,nayak,fukane,sau,potterlee,xlqi,alicea,lutchyn,yuval}. An important and promising path discovered thus far is to combine spin-orbit coupling (SOC) and Berry phase of the electrons with superconductivity. Localized Majorana zero-energy modes (MZM) have been proposed to arise in the vortex core when the Dirac fermion surface states of a topological insulator are proximity-coupled to an s-wave superconductor \cite{fukane}, or when superconductivity is induced in a semiconductor with strong Rashba SOC and time-reversal symmetry breaking Zeeman field \cite{sau}. Experimental realizations of these proposals are under active current investigations \cite{kouwenhoven,yazdani,jfjia,hzheng}. There exists, however, fundamental challenges that come with using external magnetic field induced vortices. The existence and the stability of the MZM in real materials are not guaranteed due to the low-energy vortex core states \cite{fukane,ashvin,gxu} as well as disorder and vortex creep. It is difficult, if not impossible, to move the field-induced vortex lines
individually on the Abrikosov lattice, which greatly reduces the ability to manipulate the MZM for operations such as braiding. Moreover, the requirement of external field is difficult to be integrated into quantum computation devices and limits their applications.

We propose here a new form of vortex matter - the quantum anomalous vortex matter that can support robust MZM without applying external magnetic field.
In conventional spin-singlet $s$-wave superconductors, a time-reversal symmetry breaking magnetic impurity is known to create a vortex-free defect hosting the Yu-Shiba-Rusinov (YSR) bound states \cite{yu,shiba,rusinov} inside the superconducting (SC) gap.
We find that this folklore changes in a fundamental way in $s$-wave superconductors with strong SOC. In this case, topological defect excitations can be generated by a quantized phase winding of the SC order parameter around the magnetic impurity, all without applying an external magnetic field. The role of the magnetic field is played by the combination of the exchange field and SOC as in the anomalous Hall effect. The emergence of such vortices is thus remarkably analogous to the quantum anomalous Hall effect in topological insulator thin films doped with magnetic ions \cite{xdai,qah}. Hence the term quantum anomalous vortex (QAV). We demonstrate with theoretical calculations that (i) The QAV nucleates around the magnetic ion by the exchange coupling between the local moment and spin-angular momentum locked SC quasiparticles that lowers its energy compared to the vortex-free YSR state.
(ii) When superconductivity is induced in the topological surface states, MZM emerge inside the QAV core, again without applying external magnetic field.
(iii) A remarkable property of the QAV is that the Caroli-de Gennes-Matricon (CdGM) vortex core states \cite{cdm,bardeen} with nonzero effective angular momenta are expelled into the continuum above the SC gap. A comparison of the vortex profile and core states between the QAV and the field-induced vortex is shown in Fig.~1a.  The ``gapping'' of the core states critically enhances the stability and robustness of the MZM by preventing the mixing with the CdGM states at nonzero energy \cite{fukane,ashvin,gxu}.
At low densities of magnetic ions, a new electronic matter, the QAV matter with surface MZM as depicted in Fig.~1b, would arise in such layered superconductors and provide an unprecedented platform of robust and manipulatable MZM for nonabelian quantum computing.

\begin{figure}
	\begin{center}
		\fig{3.4in}{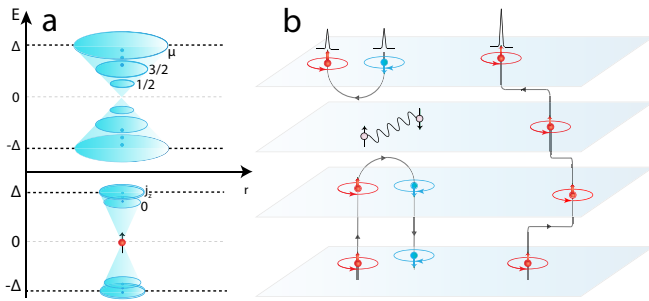}\caption{Schematic rendering of (a) a conventional magnetic field induced vortex (top panel) and the QAV nucleated at a magnetic ion (bottom panel), showing energy levels of the in-gap CdGM states localized inside the vortex core and labeled by the quantum numbers $\mu$ and $j_z$, respectively. Negative energy states are occupied. (b) Schematic rendering of the quantum anomalous vortex matter in the layered superconductor. Red and blue dots/arrows indicate opposite $c$-axis moment directions of the magnetic ions, while arrowed circles the vorticities of the QAV. Black lines are the continuous flux lines piercing SC layers through the magnetic ions. Zero-bias peaks indicate the localized MZM from the TSS in the QAV cores where the flux lines enter and leave the sample surface.}
	\end{center}
	\vskip-0.5cm
\end{figure}

We find that the QAV matter is pertinent to the Fe-based superconductor Fe(Te,Se), exhibiting spectroscopic properties remarkably consistent with the surprising discovery of robust zero-energy bound states (ZBS) near the excess Fe by STM in the {\em absence} of external magnetic fields \cite{yin}. Topological surface states (TSS) in Fe(Te,Se) \cite{zjwang,xxwu} have been observed by ARPES recently and acquire a SC gap below $T_c$ by the natural coupling to bulk superconductivity in the {\em same} crystal \cite{pzhang}. The condition for the applied magnetic field induced vortices to host ZBS is still unclear, with one group reporting its absence and CdGM vortex core states at nonzero energies \cite{hhwen} and another finding the ZBS in about 20\% of the vortices \cite{hongding}. However, the observation of ZBS at all excess Fe sites in zero-field is ubiquitous with measured properties fully consistent with MZM \cite{yin}. The excess Fe in as-grown Fe(Te,Se) are native magnetic impurities sitting at the $C_4$ symmetric interstitial site surrounded by the Te atoms responsible for the strong SOC. Neutron scattering finds that each excess Fe carries a $c$-axis local magnetic moment ($\sim2.5\mu_B$) and induces a ferromagnetic cluster in the neighboring Fe sites \cite{broholm}.
We show by explicit calculations that the ZBS are consistent with MZM localized in the QAV induced by the interstitial Fe
with the crucial caricature of the expulsion of none-zero energy CdGM states observed in the STM data \cite{yin}.

\textit{Theoretical model}
The bulk and surface electronic structures of Fe(Te,Se) have been studied in recent theoretical works \cite{zjwang,xxwu,gxu}.
In a nutshell, with increasing Te concentration, the band derived from the Te/Se $p_z$ orbital is pushed down in energy and hybridizes strongly with the Fe $d$-orbitals. The intrinsic SOC enhanced by Te then opens up a gap and induces a $p$-$d$ band inversion near the $Z$ point in the Brillouin zone, giving rise to the TSS upon projection onto the $(001)$ surface. The situation is analogous to a 3D topological insulator, with the exception that the normal state of Fe(Te,Se) is a metal and the TSS is electron doped. The bulk and the surface electronic structures have been qualitatively confirmed by the high-resolution, spin-polarized ARPES experiments recently \cite{pzhang,pzhang-new}.

We construct here an effective low-energy theory where the bands of the bulk and surface states are treated separately in an $s$-wave superconductor with intrinsic SOC. An isolated magnetic ion, such as the interstitial Fe impurity in Fe(Te,Se), is introduced at ${\bf r}=0$. Our strategy is to study first the bulk SC state and show that a topological defect excitation, where the phase of the pairing order parameter winds by $2\pi$ as in a vortex, nucleates spontaneously at the magnetic ion. This part is independent of the existence of TSS. Then, we couple such a novel QAV to the helical Dirac fermion TSS and study the emergence of the robust MZM. Using Fe(Te,Se) as a reference material, we consider a hole-like bulk band around the $\Gamma$ point as shown in Fig.~2a with the effective mass 
{$m^*\simeq4.08m_e$} and Fermi energy $\varepsilon_f\simeq-4.52$meV extracted from the ARPES experiments \cite{pzhang}. The role of the electron-like band in Fe(Te,Se) will be discussed later. In the spinor basis $\psi(\mathbf{r})=[\psi_{\uparrow}(\mathbf{r}), \psi_{\downarrow}(\mathbf{r})]^T$, the normal state Hamiltonian reads in the continuum limit,
\begin{eqnarray}
H&=&-\frac{p^2}{2m^*}-\varepsilon_f+H_{\rm soc} + H_{\rm ex}.
\label{h0}
\end{eqnarray}
Here, the SOC term is described by
\begin{equation}
H_{\rm soc}=-\lambda_{\rm so}(r){\bf L}\cdot{\boldsymbol{\sigma}},
\label{hsoc}
\end{equation}
where ${\bf L}={\mathbf r}\times{\mathbf p}$ is the angular momentum operator and $\boldsymbol{\sigma}$ the vector spin Pauli matrix. Note that in the presence of the intrinsic SOC \cite{zjwang,xxwu}, the spin rotation symmetry is broken and the band electrons carry the pseudospin quantum number. Since it will not affect our results, we use the term spin instead of pseudospin for simplicity.
The $\lambda_{\rm so}(r)$ in Eq.~(\ref{hsoc}) comes from the Elliot-Yafet SOC induced by the impurity \cite{elliot,yafet} embedded in the strongly spin-orbit coupled environment. The spin and angular momentum of the partial waves are thus locked by the impurity whose magnetic moment ${\bf I}_{\rm imp}$ involves both spin and orbital contributions of the magnetic ion. The exchange interaction in Eq.~(\ref{h0}) thus
contains both spin and orbital exchange processes  \cite{smit,kondo,Elliott-Thorpe,schrieffer,fert} and can be written in the basis of the total angular momentum ${\bf J}={\bf L}+{1\over2}{\boldsymbol{\sigma}}$ as
\begin{equation}
H_{\rm ex}=-{\cal J}_{\rm ex}(r){\bf I}_{\rm imp}\cdot{\bf J}.
\label{hex}
\end{equation}
For simplicity, the short-ranged exchange coupling ${\cal J}_{\rm ex}(r)$ is assumed to be isotropic to keep the parameters of the theory at a minimum.
Eq.~(\ref{hex}) is similar to the exchange interaction in dilute magnetic semiconductors \cite{semiconductor}.
\begin{figure*}
	\begin{center}
		\fig{7.0in}{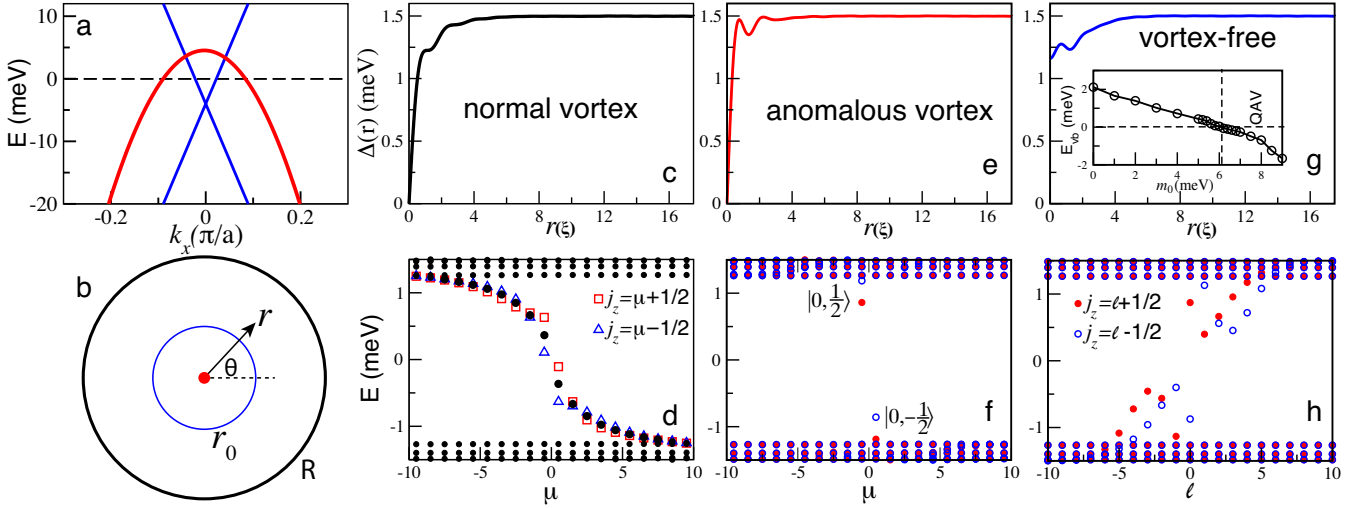}\caption{(a) The effective hole-like band for bulk states (red line) and the TSS band (blue line) observed in \cite{pzhang} near the $\Gamma$ point. (b) The disc geometry with radius $R$. Short-ranged couplings have an exponential decay-length {$r_0$} from the centered magnetic ion. (c-h) The self-consistent pairing profile $\Delta(r)$ where $r$ is measured in unit of the coherence length $\xi$ and the in-gap bound states spectrum: (c-d) The normal magnetic field induced vortex. The SOC associated with the magnetic ion with $\lambda_0=6.6$meV splits the degenerate CdGM states $\vert\mu,\sigma\rangle$ (black circles) into two sets of spin-orbit coupled bound states $\vert j_z,\sigma\rangle$ with $j_z=\mu\pm{1\over2}$ (open squares and triangles). (e-f) The QAV induced by the magnetic ion for $m_0=\lambda_0=6.6$meV. All CdGM states $\vert j_z,\sigma\rangle$ except $\vert0,\sigma\rangle$ are pushed to higher binding energies above the SC gap. (g-h) The vortex-free state near the magnetic ion. The SOC splits the in-gap YSR states into two sets. Inset in (g): the QAV binding energy as a function of the exchange field $m_0$. The QAV emerges beyond $m_0^c\simeq6.1$meV.}
	\end{center}
	\vskip-0.5cm
\end{figure*}

To study the SC state with a complex inhomogeneous Cooper pairing order parameter,
it is convenient to perform the Bogoliubov transformation
\begin{equation}
\psi_{\sigma}^\dagger(\mathbf{r})=\sum_{n}[u_{n\sigma}^{*}(\mathbf{r})
\gamma_{n}^\dagger+v_{n\sigma}^{}(\mathbf{r})\gamma_{n}],
\label{transformation}
\end{equation}
where $\gamma_n^\dagger$ creates a Bogoliubov quasiparticle. The resulting
Bogoliubov-de Gennes (BdG) equation is given by
\begin{eqnarray}
\left [\begin{array}{cc}
H & \mathbf{\Delta}(\bf r) \\
\mathbf{\Delta}^*(\bf r) & -\sigma_y  H^{*}\sigma_y
\end{array} \right] \Phi_n(\mathbf{r})=E_n\Phi_n(\mathbf{r}),
\label{mbdg}
\end{eqnarray}
where $\mathbf{\Delta}(\mathbf{r})=g\langle\psi_\downarrow(\mathbf{r})\psi_\uparrow(\mathbf{r})\rangle$ is the self-consistent pairing potential for an attraction $g$ \cite{sm}. We choose $g=11$meV such that the calculated $\vert\mathbf{\Delta}(\mathbf{r})\vert=\Delta=1.5$meV matches the bulk SC gap
determined experimentally for this band \cite{yin,pzhang}.
{The BCS coherence length is therefore $\xi=\hbar v_f/\pi\Delta\simeq2.76$nm, which is not far from the measured value $\sim2$nm in Fe(Te,Se) \cite{cdepth}.
	Diagonalizing the BdG equation yields the energy spectrum $E_n$ and Nambu wavefunctions $\Phi_n(\mathbf{r})=[u_{n\uparrow}(\mathbf{r}),u_{n\downarrow}(\mathbf{r}),v_{n\downarrow}(\mathbf{r}),-v_{n\uparrow}(\mathbf{r})]^T$ for both the vortex-free and vortex solutions with ${\mathbf\Delta}(\mathbf{r})=\Delta(r)e^{i\nu\theta}$, where the integer $\nu$ is the vorticity.
	
	We have obtained the solutions in the disc geometry for a SC layer with the isolated magnetic ion at its center (Fig.~2b) in polar coordinates ${\bf r}=(r,\theta)$. The SOC in Eq.~(\ref{hsoc}) reduces to $-\lambda_{\rm so}(r)L_z\sigma_z$ with $L_z=-i\hbar{\partial_\theta}$.
	The wavefunction is factorizable according to
	\begin{eqnarray}
	\Phi_{n\mu}(r,\theta)&=&e^{i\mu\theta}[u_{n\mu+{\nu\over2}\uparrow}(r)
	e^{i{\nu\over2}\theta},
	u_{n\mu+{\nu\over2}\downarrow}(r)e^{i{\nu\over2}\theta},
	\nonumber \\
	&&
	v_{n\mu-{\nu\over2}\downarrow}(r)e^{-i{\nu\over2}\theta},
	-v_{n\mu-{\nu\over2}\uparrow}(r)e^{-i{\nu\over2}\theta}
	]^T
	\nonumber
	\end{eqnarray}
	where the principal quantum number $n$ is determined by solving the radial $(u,v)$ in the basis of Bessel functions and the angular quantum number $\mu=\ell-{\nu\over2}$ with $\ell$ an integer \cite{cdm,bardeen,schluter,sm}. The details of the calculation are given in the supplemental material \cite{sm}.
	The calculations are performed on discs of radius $R=87.5\xi$. The SOC and exchange coupling due to the magnetic ion are assumed to decay exponentially $\lambda_{\rm so}(r)$, ${\cal J}_{\rm ex}(r)\propto e^{-{r/r_0}}$ with a common decay length $r_0=0.7\xi$ (Fig.~2b) for simplicity and easy comparison.

	\textit{Normal vortex state} In the absence of the magnetic ion,
	the self-consistent pairing profile $\Delta(r)$ for the field-induced vortex solution with $\nu=-1$ is shown in Fig.~2c. It vanishes at the vortex center, exhibits the characteristic Friedel-like oscillations \cite{schluter,machida}, and approaches the bulk value for $r>6\xi$.
	The eigenstate energies are plotted in Fig.~2d, zoomed in to the range $(-\Delta,\Delta)$ to display the in-gap CdGM vortex core states. These bound states are doubly degenerate, carry the half-integer quantum number $\mu$, and grow with $\mu$ initially as $E_\mu\propto \mu{\Delta^2\over \varepsilon_f}$, $\mu=\pm{1\over2},\pm{3\over2},\dots$ \cite{cdm,bardeen}. Owing to the small $\vert\varepsilon_f\vert$ and the large $\Delta$ in Fe(Te,Se), the CdGM states are in the quantum limit ($\xi k_f\sim2$) with the onset at $E_{-{1\over2}}\simeq0.38$meV (Fig.~2d).
	The vortex binding energy is defined as
	\begin{equation}
	E_{\rm vb}=E_{\rm vortex}-E_{\rm vortex-free},
	\label{evb}
	\end{equation}
	where $E_{\rm vortex}$ and $E_{\rm vortex-free}$ are the energy of the vortex and vortex-free state, respectively. There is an energy cost ($E_{\rm vb}>0$) for creating the normal vortex, since the supercurrent-carrying mid-gap CdGM states with $E_\mu<0$ are occupied. An external magnetic field must therefore be applied to break the time-reversal symmetry and provide the energy cost for creating the vortex cores and the magnetic flux lines in order to support such topological defect excitations in usual superconductors.
	
	Note that there exists a sense of ``chirality'' for the vortex core states shown in Fig.~2d, i.e. $\mu E_\mu<0$, which is determined by the particle/hole-vorticity ${\rm sgn}(\varepsilon_f)\nu>0$. It can be flipped by either changing the sign of $\nu$ (controlled by the direction of the external magnetic field for normal vortices) or by changing to an electron band with $\varepsilon_f>0$. As we will show below, for the QAV nucleated at the magnetic ion of a given moment polarization direction, this chirality is determined by the condition to lower the vortex binding energy with a spontaneously determined sign of the vorticity $\nu$.
	
	\textit{Quantum anomalous vortex state}
	Let's switch on the SOC due to the magnetic ion, $H_{\rm soc}=-\lambda_{\rm so}(r)L_z\sigma_z$ with $\lambda_{\rm so}(r)=\lambda_0e^{-r/r_0}$ in Eq.~(\ref{hsoc}). It splits off in energy the nonzero angular momentum partial waves with different spin-$\sigma_z$ projections \cite{sm}. While $\lambda_0$ is not known directly from current experiments, the SOC-induced band splitting in bulk FeSe and Fe(Te,Se) have been measured by ARPES to be in the range of 20-40meV \cite{borisenko,peterjohnson}. We expect $\lambda_0$ to be in a similar range. Considering the effective nature of the theory, we will use a smaller value $\lambda_0=6.6$meV of no particular significance for the electrons in the hole-like band in the rest of the paper, unless otherwise noted.
	As shown in Fig.~2d by the colored symbols, each doubly-degenerate CdGM state splits into $\vert j_z,\pm{1\over2}\rangle$ by an amount controlled by $\lambda_0$. Because $\lambda_{\rm so}(r)$ decreases exponentially, the higher orbital angular momentum $\ell$-states with wavefunctions concentrated further away from the magnetic ion experience a smaller $\lambda_{\rm so}(r)$. As a result, the SOC effect on the vortex core states is most pronounced for the small $\mu$ CdGM states. Moreover, the nonmonotonic SOC energy of the $\ell$-states leads to a sign change in the splitting of the $\vert j_z,\pm{1\over2}\rangle$ states at larger $\mu$, as can be seen in Fig.~2d.
	Since all the negative energy CdGM states are occupied, SOC alone does not significantly lower the vortex core energy.
	
	The binding energy of the QAV comes from the exchange interaction
	in Eq.~(\ref{hex}) under SOC, which is qualitatively different from the proposal of spontaneous vortex lattice in ferromagnetic superconductors \cite{varma81}. Since magnetic transition metal ions usually have a large moment, such as the excess Fe moment in Fe(Te,Se) \cite{broholm} pointing along the $c$-axis due to the magnetic anisotropy, the impurity moment can be treated classically, i.e. ${\bf I}_{\rm imp}= M{\hat z}$. The exchange interaction becomes $H_{\rm ex}=-m(r)J_z$ with
	$m(r)={\cal J}_{\rm ex}(r)M\equiv m_0e^{-r/r_0}$.
	Consequently, for $m_0>0$, the energy of the CdGM states shown in Fig.~2d is lowered (raised) by the spin-orbit exchange field $-m(r)j_z$ for all positive (negative) $j_z$. This expulsion of the in-gap CdGM vortex core states toward the continuum above the SC gap is the crucial mechanism for lowering the core energy of the QAV by the exchange field.
	For a local moment polarized in the opposite direction, the exchange field changes a sign, i.e. $m_0<0$, the vortex core energy can be lowered in the same manner if the QAV nucleates with an opposite particle/hole-vorticity ${\rm sgn}(\varepsilon_f)\nu<0$ and a flipped chirality of the vortex core states $\mu E_\mu >0$.
	
	Fig.~2e shows the self-consistent vortex profile calculated using the full Hamiltonian $H$ with a $m_0=6.6$meV.
	The vortex core sharpens considerably and the Friedel oscillations become more prominent compared to the normal field-induced vortex in Fig.~2c.
	The eigenstate energies
	are plotted in Fig.~2f. All but two of the CdGM states $\vert j_z,\sigma_z\rangle$ are expelled out of the gap center
	into the continuum. The remaining two sitting just below the gap edges have $j_z=0$ and are thus unaffected by the exchange field. They could have been pushed further toward the continuum by a larger $\lambda_0$. The enhanced binding energy of the occupied CdGM states significantly lowers the energy of the anomalous vortex state.
	In contrast, the vortex-free ($\nu=0$) state obtained self-consistently using the same parameters shows a much broader deformation in $\Delta(r)$ near the magnetic ion (Fig.~2g) and fosters two sets of spin-orbit coupled mid-gap YSR states (Fig.~2h) that must be occupied at an energy cost.
	Thus, the energy of the quantum anomalous vortex state nucleated at the magnetic ion can be lower than that of the vortex-free state, provided that the exchange field $m_0$ is sufficient to drive $E_{\rm vb}<0$.
	
	We next calculate the vortex binding energy as a function of $m_0$, which is shown in the inset of Fig.~2g. $E_{\rm vb}$ decreases approximately linearly with increasing $m_0$ and a transition from the vortex-free YSR state to the QAV state occurs at a critical $m_0^c\simeq6.1$meV. For $m_0>m_0^c$, $E_{\rm vb}<0$, and it becomes more energetically favorable for the SC order parameter to develop the quantized phase winding with supercurrents flowing around the magnetic ion. Hence the formation of the QAV. We have studied the binding energy for different values of $\lambda_0$. It turns out that the critical exchange field is only weakly dependent on $\lambda_0$. For example, for $\lambda_0=10$meV, $m_0^c\simeq5.85$meV. The small decrease comes from the increased binding energy of the $j_z=0$, $\vert 0,\pm1/2\rangle$ states in Fig.~2f, as they are located closer to the continuum due to a larger $\lambda_0$.
	In Ref.~\cite{broholm}, it was shown that the local magnetic structure induced by the interstitial Fe as observed in the neutron scattering experiments, which involves $\sim50$ lattice Fe atoms, can be described by a five-orbital Hubbard model with a magnetic ion induced spin exchange interaction on the order of $70$meV. The intrinsic magnetic correlations in Fe(Te,Se) can also renormalize the magnitude and the extent of the exchange fields induced by the magnetic impurity ion. While how the latter translate into the effective continuum theory remains to be investigated, we will regard $m_0$ as a phenomenological parameter that controls the binding energy of the QAV.
	
	It is instructive to estimate the line energy (tension) of the QAV, which is given by \cite{deGennesbook} ${\cal E}_{\rm line}=(\phi_0/4\pi\lambda_p)^2\ln\kappa$, where $\phi_0=hc/2e$ is the SC flux quantum, $\lambda_p=(m^*c^2/4\pi n_s e^2)^{1/2}$ is the penetration depth, and $\kappa=\lambda_p/\xi$. It can be written in the form
	\begin{equation}
	{\cal E}_{\rm line}={1\over3\pi^2}{\varepsilon_f^2\over\Delta}{1\over\xi}\ln\kappa\simeq {0.46\over\xi}\ln\kappa\ {\rm meV},
	\end{equation}
	where $\ln\kappa$ is typically of order one for type-II superconductors. For Fe(Te,Se), the measured coherence length $\sim2$nm and penetration depth $\sim500$nm \cite{cdepth,pdepth1,pdepth2,pdepth3} provide an estimate ${\cal E}_{\rm line}\sim2.5{\rm meV}/\xi$, i.e. it costs on the order of $2.5$meV for a straight vortex line nucleated at an isolated magnetic ion to extend over a coherence length across the SC layers. Thus, for moderate QAV binding energies, a straight vortex line can only penetrate a few SC layers without encountering a ``boost'' by another interstitial Fe. However, the vortex lines can travel between the layers, taking advantage of the smaller line energy, and pierce through the SC planes where magnetic ions reside via the nucleation of QAVs as shown schematically in Fig.~1b.
	
	To develop further insights, we elucidate qualitatively the condition for the emergence of the QAV analytically. The lowering of the core energy due to the orbital exchange field is $E_1=\int d^2r \sum_\sigma m(r) \psi_\sigma^\dagger(r) L_z\psi_\sigma(r)/\hbar$. For an $s$-wave superconductor, the total angular momentum is quantized and given by $\hbar N/2$ in a vortex state, where $N/2$ is the number of Cooper pairs \cite{volovik95,nygaard03,tada15,Prem17}. We thus obtain $E_1\simeq\pi m_0r_0^2 n^{2D}$, where $n^{2D}=nd$ is the 2D particle density and $d$ the layer thickness. This is to be compared to the energy cost of the vortex line in the 2D layer $E_2=d{\cal E}_{\rm line}\simeq {\pi\hbar^2\over4m^*}n_s^{2D}\ln\kappa$, where $n_s^{2D}$ is the 2D superfluid density. Thus, at zero temperature, the QAV nucleates when $E_1>E_2$, i.e.
	\begin{equation}
	m_0> {\hbar^2\over4m^*r_0^2}\ln\kappa={\vert \varepsilon_f\vert\over2(r_0k_f)^2}\ln\kappa.
	\label{m0}
	\end{equation}
	For short-range exchange interactions $r_0\sim1/k_f$,
	the critical exchange field $m_0^c$ is therefore on the order of the Fermi energy $\vert \varepsilon_f\vert$, consistent with the numerical result shown above ($r_0 k_f\simeq1.4$ for our parameters). In this sense, the small Fermi energy and superfluid density/stiffness \cite{rhos} in Fe(Te,Se) favor the emergence of the QAV state. On the other hand, if the decay length $r_0\sim\xi$, Eq.~(\ref{m0}) implies $m_0^c\simeq{\pi^2\over8}{\Delta^2\over \vert \varepsilon_f\vert}\ln\kappa$, which is on the scale of the CdGM mini-gap energy and also very favorable for the nucleation of the QAV. While for superconductors in the quantum limit where $k_f\xi\sim1$, such as Fe(Te,Se), these two limits are essentially equivalent, it is noteworthy that even for superconductors with a substantial Fermi energy, the QAV can still be induced by a reasonable exchange interaction provided that its decay length in the low-energy effective theory is on the order of the SC coherence length, which can originate from the underlying magnetic correlations in the superconductor \cite{khaliullin} or in the presence of sizable magnetic quantum dots.
	
	\textit{Multiband and composite QAV} By focusing on a single hole-like band around the $\Gamma$ point in the above discussion, we have studied the physical origin and the basic properties of the QAV. We now discuss the role of the electron bands in Fe(Te,Se) and demonstrate that it leads to a nontrivial test and brings out new physics of the QAV in multiband superconductors. For simplicity, we ignore the weak interband Josephson coupling and consider a single isotropic electron band in the continuum limit with band mass $m_e^*\simeq1.33m_e$, Fermi energy $\varepsilon_f^e\simeq25$meV, and pairing gap $\Delta_e\simeq4$meV extracted from the ARPES experiments \cite{kunjiang,pzhang-prb,pzhang-new,miaohu}. The crucial observation is that for a fixed polarization of the magnetic ion with $m_0>0$ as before, the QAV nucleated from the electron band must preserve the chirality $\mu E_\mu<0$ of the vortex core states in order for the exchange field to lower its energy by pushing the CdGM states into the continuum. This requires an unchanged particle/hole-vorticity ${\rm sgn}(\varepsilon_f^e)\nu_e>0$ for the anomalous vortex from the electron band. As a result, the energetically favorable QAV from the electron band must originate from the corresponding order parameter ${\mathbf\Delta}_e(\mathbf{r})=\Delta_e(r)e^{i\nu_e\theta}$ with an opposite phase winding, i.e. an opposite vorticity $\nu_e=1$ compared to the hole band.
	
	In the supplemental material \cite{sm}, we provide the details of the calculation of the QAV binding energy as a function of $m_0>0$ for the electron band. Since the impurity induced SOC $\lambda_{\rm so}$ is inversely proportional to the band mass both in sign and in magnitude \cite{elliot,yafet}, we thus use $\lambda_0^e=-21$meV for the electron band correspondingly, while keeping $r_0$ the same. The results show that the binding energy has an approximately linear dependence on $m_0$ similar to the hole-band case shown in the inset of Fig.~2g, leading to a critical $m_0^{c,e}\simeq23.2$meV beyond which the QAV emerges, consistent with the analytical expression discussed above. Thus, for a reasonably strong exchange field, composite QAVs can nucleate at the excess Fe sites where the supercurrents are carried by electrons from both the hole and electron bands in Fe(Te,Se). Such composite QAVs can be remarkable realizations of the composite vortices with fractional flux proposed and studied both theoretically and experimentally in multiband superconductors \cite{babaev02,bluhm06,lin13}.
	
	\begin{figure}
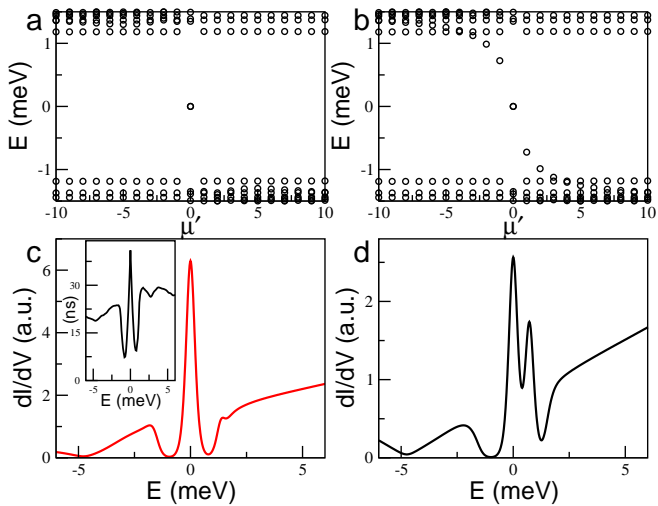

		\begin{center}
			\fig{3.4in}{fig3.eps}\caption{The low-energy spectrum of the TSS coupled to (a) the QAV induced by the magnetic ion (see Figs 2e and 2f) showing an isolated MZM, and (b) a conventional magnetic field induced vortex without the magnetic ion. The parameters used for the TSS are $\lambda_0^\prime=m_0^\prime=6.6$meV.
				(c) The calculated tunneling conductance (local density of states) at the QAV center. Inset: STM tunneling conductance at the excess Fe site in Fe(Te,Se) \cite{yin}. (d) The tunneling conductance obtained at the center of the magnetic field induced vortex in (b), showing multiple CdGM vortex core states. Calculated conductance spectra are broadened by a temperature $T=1.5$K.
			}
		\end{center}
		\vskip-0.5cm
	\end{figure}
	
	\textit{Majorana zero-energy bound state in QAV} We turn to the emergence of localized MZM in the QAV when superconductivity is induced in the helical Dirac fermion TSS. In the vicinity of the magnetic ion, the effective Hamiltonian in the continuum limit, with primes indicating for the TSS,
	can be written in the spinor basis as,
	\begin{eqnarray}
	H^\prime &=&v_D({\boldsymbol{\sigma}}\times\mathbf{p})\cdot\mathbf{z}
	- \varepsilon_f^\prime + H_{\rm soc}^\prime + H_{\rm ex}^\prime,
	\label{dh0}
	\end{eqnarray}
	where the velocity $v_D=0.216$eV$\cdot\mathring{A}$ and the Fermi level $\varepsilon_f^\prime=4.5$meV above the Dirac point of the electron doped TSS band were extracted from the ARPES experiment on Fe(Te,Se) \cite{pzhang}, as shown in Fig.~2a.
	The impurity-induced SOC is $H_{\rm soc}^\prime=\lambda_{\rm so}^\prime(r)L_z\sigma_z$ and $H_{\rm ex}^\prime=-{\cal J}_{\rm ex}^\prime(r)I_{\rm imp}^z J_z$ is the exchange coupling between the TSS and the local moment \cite{sczhang-ex}. In general, $\lambda_{\rm so}^\prime(r)$ and ${\cal J}_{\rm ex}^\prime(r)$ can be different from those in the bulk states. We consider here the QAV formed in the hole band alone. In the corresponding BdG equation (\ref{mbdg}), the induced pairing potential for the TSS is $\mathbf{\Delta}^\prime(\mathbf{r})=\Delta_{\rm QAV}(r)e^{i\theta}$, where $\Delta_{\rm QAV}(r)$ is the pairing profile of the QAV shown in Fig.~2e.
	The obtained vortex core states energy spectrum for the TSS is plotted in Fig.~3a. The ``isolated'' bound state at $E=0$ is precisely the MZM, i.e. the $\mu^\prime=0$ element of the chiral CdGM states $E_{\mu^\prime}\propto \mu^\prime\Delta^2/\varepsilon_f^\prime$, where $\mu^\prime=0,\pm 1,\pm2,\cdots$ is now an integer due to the additional Berry phase of the Dirac fermions \cite{ashvin,sm}.
	Note that all other CdGM states with nonzero $\mu^\prime$ in Fig.~3b obtained without coupling to the magnetic ion are pushed into the continuum above the SC gap by the exchange field
	via the same mechanism that produced the QAV. The gapping of the nonzero energy CdGM states prevents the level crossing induced topological vortex transition \cite{ashvin} and protects the robustness of the MZM even for higher doping levels of the TSS.
	
	There are remarkable agreements between the calculated local density of states at the center of the QAV plotted in Fig.~3c and the tunneling conductance measured by STM \cite{yin} at the interstitial excess Fe sites reproduced in the inset. Both show the V-dip around $-4.5$meV corresponding to the Dirac point of the TSS, the absence of coherence peaks at the gap energies $\pm\Delta$, and a spectrum free of mid-gap states other than the zero-bias peak. Note that without coupling to the magnetic ion, the mid-gap vortex core states produce multiple conductance peaks and reduce considerably the spectral weight of the MZM as shown in Fig.~3d. Indeed, the small satellite peak at $\sim1.44$meV in Fig.~3c comes from the CdGM states expelled into the continuum by the exchange field. It is tempting to identify a similar satellite in the STM spectrum at $\sim1.6$meV with such a resonance. These findings further support that the ZBS observed by STM are MZM in the QAV induced by the excess Fe. Since the QAV has trapped a SC flux quantum, flux-quantization as well as the nature of the MZM protect the QAV and the zero-energy bound state from external magnetic fields applied along the $c$-axis, consistent with the insensitivity of the ZBS to fields up to $8$ Tesla observed by STM \cite{yin}.
	
	\textit{Quantum anomalous vortex matter}
	Let's consider a low density of dilute magnetic ions such that the SC transition temperature $T_c$ is reduced but the SC ground state remains stable.
	For Fe$_{1+y}$Te$_{0.55}$Se$_{0.45}$, this is the case for excess Fe density $y<0.03$ \cite{pan}. The theory thus predicts a QAV matter illustrated in Fig.~1b with localized MZM at the ends of the flux lines as they enter or leave the sample's surface via the magnetic ion. Note that while the vorticity of the QAV is confined to the magnetic ion, the magnetic flux lines must be continuous. Fig.~1b illustrates the case where the magnetic flux lines pierce through the SC layers where magnetic ions reside by the nucleation of a QAV. An immediate consequence is the appearance of QAV with opposite vorticities, i.e. both vortices and anti-vortices accompanied by the flipping of the local magnetic moment direction along the $c$-axis. This prediction could serve as an experimental test of the theory by spin-polarized STM and/or scanning probe SQUID.
	
	The QAV matter is likely to have been realized in Fe(Te,Se) superconductors. A crucial observation of Ref.\cite{yin} is that when two interstitial Fe atoms sit close together, a striking simultaneous reduction in the amplitudes of the zero-bias peaks occurs without observable energy shifts. As shown in Fig.~1b, the continuous magnetic flux line now favors the formation of an entangled quantum anomalous vortex-antivortex pair, providing a natural explanation via the annihilation of a pair of MZM. It remains to be understood, however, why such annihilation would not cause detectable splitting of the
	MZM from zero energy. With increasing excess Fe concentration, the density of the QAV increases, which may provide a novel mechanism for the suppression of bulk superconductivity and the eventual superconductor to metal quantum phase transition.
	
	Combining the QAV matter with either SC topological surface states such as in Fe(Te,Se) or a topological superconductor provide advantageous platforms in zero external magnetic fields for studying the statistics and interactions of the MZMs. In order to accomplish quantum braiding in such systems, two conditions, which are in principle achievable but undoubtedly challenging, need to be satisfied. First, the STM tip should only ``grab and drag'' the magnetic impurity ions but not the atoms in the underlying lattice to avoid drastic changes to the local electronic states. Second, the entire braiding process should be done as slowly as possible to satisfy the adiabatic condition. As long as the manipulations are in the adiabatic limit, the low energy Majorana bound state description will be valid and the nonabelian statistics of the MZM can be realized. Furthermore, if the STM is used for both {\em manipulating} the excess Fe atoms (and the associated QAVs) and {\em measuring} the information in the MZMs, the tunneling current will introduce a source of dissipation and dephasing that put an upper bound for the adiabatic braiding time, accounting also for the extrinsic dissipation caused by the environment. This is also true when using techniques other than the STM, as well as using other platforms such as the Majorana nanowires \cite{karsten}. Thus, in order to braid two MZMs successfully before the loss of coherence, the current associated with the STM needs to be sufficiently weak. An alternative approach is to separate the manipulation from the measurement processes. The MZMs can be positioned in designed arrays by the STM. Then, using the algorithm of braiding in measurement-only \cite{bonderson}, the adiabatic and coherence time constraints can be satisfied more favorably by anyonic interferometry measurements on the designed arrays.

	We thank Shuheng Pan, Jiaxin Yin, Andrew Potter, Hong Ding, Jiangping Hu, and Tao Xiang for valuable discussions. This work is supported by the U.S. Department of Energy, Basic Energy Sciences Grant No. DE-FG02-99ER45747 (K.J, Z.W). K.J. also thank the hospitality of Institute of Physics, Chinese Academy of Sciences and the support of the Ministry of Science and Technology of China 973 program (No. 2017YFA0303100, No. 2015CB921300). X. D. acknowledges the financial support from the Hong Kong Research Grants Council (Project No. GRF16300918).

\newpage
\renewcommand{\theequation}{S\arabic{equation}}
\renewcommand{\thefigure}{S\arabic{figure}}
\renewcommand{\thetable}{S\arabic{table}}
\setcounter{equation}{0}
\setcounter{figure}{0}

\title{Supplemental material for nodeless high-T$_c$ superconductivity in highly-overdoped monolayer CuO$_2$}

\maketitle
\begin{widetext}
\section*{Supplemental Material}


\section{A. BdG equation and vortex solution}
We provide more details of the calculations for the emergence of a QAV and MZM at an isolated interstitial magnetic ion in s-wave superconductors with strong SOC in the absence of applied external magnetic fields. The Hamiltonian of the system has been discussed in the main text for both the bulk states and the TSS. For convenience, we write the total Hamiltonian as $H=H_{n}+H_p$, where $H_n$ is the normal part and $H_p$ the pairing part. They are given by,
\begin{eqnarray}
H_n&=& H_{\rm kin}+H_{\rm soc}+H_{\rm ex}
\label{normalh} \\
H_p&=&\int d\mathbf{r}\mathbf{\Delta(r)}\psi_\uparrow^{\dagger}(\mathbf{r})
\psi_\downarrow^{\dagger}(\mathbf{r})+h.c..
\label{pairingh}
\end{eqnarray}
The complex pairing potential $\mathbf{\Delta}(r)=g\langle\psi_\downarrow(\mathbf{r})
\psi_\uparrow(\mathbf{r})\rangle$ is determined self-consistently for an attraction $g$. In Eq.~(\ref{normalh}), $H_{\rm soc}$ and $H_{\rm ex}$ are the SOC and the exchange interaction produced by the magnetic ion discussed in their operator forms in Eqs (2) and (3) in the main text, while $H_{\rm kin}$ describes the different kinetic energy of the bulk band and the surface states in our effective theory. The gauge invariance requires the use of canonical momentum operators in the Hamiltonian, i.e. ${\bf p}\to {\boldsymbol{\pi}}={\bf p}-{e\over c}{\bf A}$ where ${\bf A}(\mathbf{r})$ is the vector potential. In the spinor notation, $\psi(\mathbf{r})=(\psi_{\uparrow}(\mathbf{r}),
\psi_{\downarrow}(\mathbf{r}))^T$,
\begin{eqnarray}
H_{kin}=\int d\mathbf{r}\psi^{\dagger}(\mathbf{r})
\left[-{1\over2m^*}({\bf p}-{e\over c}{\bf A})^2-\varepsilon_f\right]\psi(\mathbf{r})
\label{hkin-bulk}
\end{eqnarray}
describes the parabolic dispersion of a hole-like bulk band near ${\bf p}=0$ ($\Gamma$ point) in the continuum limit, and
\begin{eqnarray}
H_{kin}^\prime=\int d\mathbf{r}\psi^{\dagger}(\mathbf{r}) \left[v_D({\boldsymbol{\sigma}}\times\boldsymbol{\pi})\cdot\mathbf{z}-
\varepsilon_f'\right]\psi (\mathbf{r})
\label{hkin-tss}
\end{eqnarray}
the helical Dirac fermion TSS. As in the main text, we will continue to use primed quantities for the TSS.
The total Hamiltonian $H$ can be solved conveniently using the Bogoliubov transformation
\begin{eqnarray}
\psi_{\sigma}(\mathbf{r})=\sum_{n}\bigl[u_{n\sigma}(\mathbf{r})\gamma_{n}
+v_{n\sigma}^{*}(\mathbf{r})\gamma_{n}^\dagger\bigr], \qquad
\psi_{\sigma}^\dagger(\mathbf{r})=\sum_{n}\bigl[u_{n\sigma}^{*}
(\mathbf{r})\gamma_{n}^\dagger+v_{n\sigma}(\mathbf{r})\gamma_{n}\bigr], \label{phi}
\end{eqnarray}
where $\gamma_n^\dagger$ and $\gamma_n$ are the creation and destruction operators of a Bogoliubov quasiparticle,
\begin{eqnarray}
\gamma_{n}^\dagger=\int d\mathbf{r}\sum_{\sigma}\bigl[u_{n\sigma}(\mathbf{r})\psi_{\sigma}^{\dagger}
(\mathbf{r})+v_{n\sigma}(\mathbf{r})\psi_{\sigma}(\mathbf{r})\bigr], \quad
\gamma_{n}=\int d\mathbf{r}\sum_{\sigma}\bigl[u_{n\sigma}^{*}(\mathbf{r})\psi_{\sigma}
(\mathbf{r})+v_{n\sigma}^{*}(\mathbf{r})
\psi_{\sigma}^\dagger(\mathbf{r})\bigr].
\label{gamma}
\end{eqnarray}
In terms of the Nambu spinors $\Phi_n(\mathbf{r})=(u_{n\uparrow}(\mathbf{r}),u_{n\downarrow}(\mathbf{r}),
v_{n\downarrow}(\mathbf{r}),-v_{n\uparrow}(\mathbf{r}))^T$, the Schr\"odinger equation can be written as a BdG equation,
\begin{eqnarray}
\left [\begin{array}{cc}
\widehat H_n({\bf A}) & \mathbf{\Delta(r)} \\
\mathbf{\Delta}^*(\bf r) & -\sigma_y \widehat H_n^{*}({\bf A})\sigma_y
\end{array} \right] \Phi_n(\mathbf{r})=E_n\Phi_n(\mathbf{r}),
\label{bdg}
\end{eqnarray}
where $\widehat H_n$ is the operator corresponding to the normal part of the Hamiltonian in Eq.~(\ref{normalh}).
We studied both vortex-free and vortex solutions with the pairing potential ${\mathbf\Delta}(\mathbf{r})=\Delta(r)e^{i\nu\theta}$, where the integer winding number $\nu$ is the vorticity. In each case, the BdG equation is diagonalized to obtain the quasiparticle energy spectrum $E_n$ and the eigenstate wavefunctions $\Phi_n(r)$. The gap function is then calculated as in the standard BCS theory,
\begin{eqnarray}
\mathbf{\Delta}(\mathbf{r})=\frac{g}{2}\sum_{E_n\le\omega_D}\bigl[u_{n\uparrow}
(\mathbf{r})v_{n\downarrow}^*(\mathbf{r})-u_{n\downarrow}
(\mathbf{r})v_{n\uparrow}^*(\mathbf{r})\bigr],
\label{gap}
\end{eqnarray}
where $g$ is the attraction and $\omega_D$ is the energy cutoff. 
Concurrently, the spatially varying current density is determined using \cite{Sgygi}
\begin{eqnarray}
{\bf j}({\bf r})&&={e\hbar\over2m^*i}\sum_{n\sigma}\biggl[v_{n\sigma}(\mathbf{r})\bigl(\boldsymbol{\nabla}-{ie\over \hbar c}{\bf A}\bigr) v_{n\sigma}^*(\mathbf{r})-h.c.\biggr]+{e\over2}\sum_{n\sigma}
\biggl[v_{n\sigma}(\mathbf{r})\bigl(-r\lambda_{\rm so}(r)\bigr) v_{n\sigma}^*(\mathbf{r})+h.c.\biggr]{\hat{\boldsymbol\theta}}.
\label{current}
\end{eqnarray}
Note that although a QAV is obtained in the absence of the external magnetic field (i.e. without the external vector potential), the circulating current ${\bf j}(\mathbf{r})$ will generate a dynamic vector potential according to the Maxwell equation $\boldsymbol{\nabla}\times\boldsymbol{\nabla}\times{\bf A}(\mathbf{r})={4\pi\over c}{\bf j}(\mathbf{r})$, from which the profile of ${\bf A}(\mathbf{r})$ can be obtained. This procedure can be repeated by inserting the calculated $\mathbf{\Delta}(\mathbf{r})$ and ${\bf A}(\mathbf{r})$ back into the BdG equation (\ref{bdg}) until self-consistency is reached \cite{Sgygi}.

Let's first consider the case of the parabolic bulk band. The case of the Dirac fermion TSS will be discussed later. In order to determine the quantum numbers of the vortex states, it is convenient to transform to the London gauge \cite{Scdm} where the pairing potential ${\mathbf\Delta}(\mathbf{r})$ is real and the quasiparticle wavefunctions have well defined properties under a $2\pi$ rotation \cite{SdGbook}. This is achieved by the following transformation $\Phi_n(\mathbf{r})\to\Psi_n(\mathbf{r})=e^{-i{\nu\over2}\theta\tau_z}\Phi_n(\mathbf{r})$, where $\tau_z=\pm1$ acts in the particle-hole channel.
The covariant paring potential $\mathbf{\Delta}(\mathbf{r})\to \mathbf{\Delta}^\prime(\mathbf{r})=\mathbf{\Delta}(\mathbf{r})e^{-i\nu\theta}=\Delta(\mathbf{r})$ is real and the BdG equation becomes,
\begin{eqnarray}
\left [\begin{array}{cc}
\widehat H_n({\bf A}^\prime) & {\Delta(r)} \\
{\Delta(r)} & -\sigma_y \widehat H_n^{*}({\bf A}^\prime)\sigma_y
\end{array} \right] \Psi_n(\mathbf{r})=E_n\Psi_n(\mathbf{r}),
\label{bdg-1}
\end{eqnarray}
where the transformed vector potential
\begin{equation}
{\bf A}^\prime(\mathbf{r})={\bf A}(\mathbf{r})-{\nu\hbar c\over2er}{\hat{\boldsymbol\theta}}.
\label{gaugefield}
\end{equation}
To utilize the rotational symmetry about the $z$-axis, we study a SC layer in the disc geometry shown in Fig.~2b, with an isolated magnetic ion located at the center in polar coordinates ${\bf r}=(r,\theta)$. Since our low-energy effective theory separately treats the bulk band and the TSS, we ignore the dispersion along the $z$-direction for simplicity \cite{Snote}. In this gauge, the quasiparticle wavefunction $\Psi_n$ acquires a multiplicative factor of $(-1)^\nu$ under a $2\pi$ rotation \cite{SdGbook}, since the vector potential in Eq.~(\ref{gaugefield}) produces a magnetic flux line through the center of the vortex that carries $\nu$ number of SC flux quantum. Consequently, when $\Psi_n(\mathbf{r})$ is expanded into partial waves, i.e.
$\Psi_n(\mathbf{r})=e^{i\mu\theta}\Psi_{n\mu}(r)$, we obtain $\mu=\ell-{\nu\over2}$ where $\ell$ is an integer. As a result, the quantum number $\mu=\pm{1\over2},\pm{3\over2},\dots$ is a half-odd integer for the vortex states when $\nu$ is odd, i.e. for vortices of odd vorticity. Note that in the original paper of Caroli et. al. \cite{Scdm}, an error was made with respect to the property of $\mu$, which was corrected later by de Gennes \cite{SdGbook}. Substituting $\Psi_n(\mathbf{r})$ into the BdG equation (\ref{bdg-1}), the kinetic energy terms read
\begin{equation}
\tau_z\bigl[-i\hbar\boldsymbol{\nabla} - \tau_z{e\over c}{\bf A}^\prime(\mathbf{r})\bigr]^2e^{i\mu\theta}\Psi_{n\mu}(r)
=e^{i\mu\theta}\tau_z\bigl[-i\hbar\boldsymbol{\nabla}-\tau_z{e\over c}{\bf A}(\mathbf{r})+(\mu+\tau_z{\nu\over2}){1\over r}\bigr]^2\Psi_{n\mu}(r).
\end{equation}
Having determined the quantum number $\mu$ of the vortex states, it is clear that we could have started with the BdG equation (\ref{bdg}) and make the following change of variables
$$\Phi_n(\mathbf{r})=e^{i\mu\theta+i{\nu\over2}\tau_z\theta}\Psi_{n\mu}(r)$$
to arrive at the correct wavefunction \cite{Sbardeen,Sgygi}. Written out explicitly,
\begin{eqnarray}
\Phi_{n\mu}(r,\theta)=e^{i\mu\theta}[u_{n\mu+{\nu\over2}\uparrow}(r)
e^{i{\nu\over2}\theta},
u_{n\mu+{\nu\over2}\downarrow}(r)e^{i{\nu\over2}\theta},
v_{n\mu-{\nu\over2}\downarrow}(r)e^{-i{\nu\over2}\theta},
-v_{n\mu-{\nu\over2}\uparrow}(r)e^{-i{\nu\over2}\theta}
]^T,
\label{phi-mu}
\end{eqnarray}
where the principal quantum number $n$ is determined by solving the radial wavefunctions $u(r),v(r)$ in the resulting BdG equation. Nevertheless, Eq.~(\ref{gaugefield}) uncovers an important point.
In the regime $r\ll\lambda_p$ with $\lambda_p$ the penetration depth, it is known that the vector potential $A_\theta(r)\sim {1\over2}r h_{\rm eff}$ \cite{Scdm,Sbardeen}, where $h_{\rm eff}={\nu\phi_0\over2\pi\lambda_p^2}$ is the effective field along the $z$-direction and $\phi_0={hc\over2e}$ is the SC flux quantum. It reaches a maximum around $r\sim\xi$ where $\xi$ is the coherence length \cite{Sgygi}. Thus, the ratio of the vector potential to the gauge field is bounded by the order of $(\xi/\lambda_p)^2$. As a result, the effects of the vector potential ${\bf A}(\mathbf{r})$ and the magnetic field are very small and negligible for type-II superconductors where $\xi\ll\lambda_p$. As it is common practice \cite{Scdm,Sgygi90,Smachida}, we ignore ${\bf A}(\mathbf{r})$ in our calculations for simplicity. In the regime $r\gg\lambda_p$, ${\bf A}^\prime(\mathbf{r})\to0$. The main effect of the self-consistent vector potential is to screen out the supercurrent outside the vortex and cutoff the vortex line energy $\sim\rho_s\ln(\lambda_p/\xi)$ by the penetration depth $\lambda_p$, where $\rho_s$ is the superfluid density/stiffness. This is approximately accounted for by considering a disc of radius $R$ under the open boundary condition $\Delta(r)=0$ for $r>R\sim\lambda_p$. Thus, our calculated vortex binding energy and the transition from the vortex-free to the QAV phase are good estimates that can be considered as upper bounds.
{By fitting the ARPES data around the $\Gamma$ point in FeSe and Fe(Te,Se) superconductors \cite{Skunjiang,Spzhang,Spzhang-new}, we construct a parabolic hole-like band in the continuum limit with an effective mass $m^*\simeq4.08m_e$ and a Fermi energy $\varepsilon_f\simeq-4.52$meV. The BCS coherence length is $\xi=\hbar v_f/\pi\Delta\approx2.76$nm for the parabolic dispersion considered, which roughly agrees with the experimental value of $\sim2$nm for the coherence length \cite{Scdepth}.
	The low superfluid density $\rho_s$ in Fe(Te,Se) superconductors \cite{Srhos}, which is considerably smaller than even the cuprate superconductors, ensures a small vortex line energy. The line tension of the QAV is estimated in the main text. Here we give an estimate of the magnitude of the magnetic field $H(0)$ at the center of the anomalous vortex. The latter is given by twice the value of $H_{c1}$ \cite{SdGbook}, i.e. $H(0)={\phi_0\over2\pi\lambda_p^2}\ln\kappa$. Using the experimental values of $\lambda_p$ and $\xi$, we find $H(0)\sim70$G. This value becomes much smaller ($\sim2$G) if the estimate is done using only the single hole-like band in the effective theory.
	In either case, the magnetic field is very weak and its Zeeman energy can be ignored, especially compared to the exchange field $m_0$ already present due to the magnetic ion. For numerical convenience, we define a length $l_0$ such that $\frac{\hbar^2}{2m^*\l_0^2}=10$meV, which gives $l_0=0.966nm\approx0.35\xi$. Setting $l_0\equiv1$, all lengths are dimensionless in unit of $l_0$. The numerical results reported here are obtained on discs of radius $R=250$. The magnetic ion induced SOC and exchange coupling are modeled with an exponential decay length $r_0=2$. We choose $g=11$meV and $\omega_D=4.7$meV such that the pairing gap approaches the bulk value $\Delta=1.5$meV far away from the magnetic ion.}

Following the pioneering works of Caroli, de Gennes, and Matricon \cite{Scdm}, and Bardeen et. al \cite{Sbardeen}, the $r$-dependent radial functions can be conveniently expanded in the basis of the Bessel functions \cite{Sgygi,Smachida}. The self-consistent pairing function profiles $\Delta(r)$ are then plotted in Figs~2c,e,g for the three cases studied and the corresponding energy level spectra are shown in Figs~2d,f,h. To compare to the tunneling conductance measured by STM, we calculate the local density of states (LDOS) as a function of bias energy according to
\begin{eqnarray}	
\frac{dI}{dV}(r,V)\propto\sum_{n\sigma}\left[|u_{n\sigma}(r)|^2f'(E_n-eV)
+|v_{n\sigma}(r)|^2f'(E_n+eV)\right], \nonumber
\end{eqnarray}
where $f'(E)$ is the derivative of the Fermi distribution function $f(E)$. We include a thermal broadening with a temperature $T=1.5$K in the calculated LDOS of the TSS at the magnetic ion site plotted in Fig.~3c and 3d, which is the lowest temperature at which the STM tunneling conductance is measured.

\section{B. Vortex-Free and Vortex Solutions for Bulk States}
\paragraph*{Vortex-free solutions}
For the bulk vortex-free YSR state, $\nu=0$ and $\mathbf{\Delta(r)} =\Delta(r)$. The wavefunction in Eq.~(\ref{phi-mu}) becomes
\begin{eqnarray}
\Phi_{n\ell}(\mathbf{r})=e^{i\ell\theta}[u_{n\ell\uparrow}(r),
u_{n\ell\downarrow}(r),v_{n\ell\downarrow}(r),-v_{n\ell\uparrow}(r)]^T. \nonumber
\end{eqnarray}
The BdG equation (\ref{bdg}) is solved in the subspace of fixed angular momentum quantum number $\ell$ by projecting the radial wavefunctions $u_{n\ell}(r)$ and $v_{n\ell}(r)$ onto a set of Bessel functions normalized on the disc,
\begin{eqnarray}
\bigl[u_{n\ell\sigma}(r),v_{n\ell\sigma}(r)\bigr]
&=&\sum_{j=1}^N \bigl[u_{nj\ell\sigma},v_{nj\ell\sigma}\bigr]\phi_{j\ell}(r),
\end{eqnarray}
where
\begin{eqnarray}
\phi_{j\ell}(r)&=&\frac{\sqrt{2}}{RJ_{\ell+1}(\beta_{j\ell})}J_\ell
\left(\beta_{j\ell}\frac{r}{R}\right),\qquad j=1,...,N.
\label{bessel}
\end{eqnarray}
Here, the argument $\beta_{j\ell}$ is the $j$-th zero of the $\ell$-th order Bessel function of the first kind $J_\ell(x)$. Since there is an infinite number of zeros for $J_\ell(x)$, $N$ is introduced as a cutoff for $j$, which determines the dimension of the BdG equation in each $\ell$-channel. We also choose a cutoff $L_c$ for the highest angular momentum channel to be considered. The BdG equation thus reduces to a $4N\times4N$ matrix eigenvalue problem
\begin{eqnarray}
\begin{bmatrix}
T_\ell-L_\ell-M_\ell-\Lambda_\ell & 0& \Delta_\ell &0 \\
0 & T_{\ell}-L_\ell+M_\ell+\Lambda_\ell & 0 & \Delta_{\ell} \\
\Delta_\ell^T &0 & -T_{\ell}-L_\ell-M_\ell+\Lambda_\ell & 0  \\
0 & \Delta_{\ell}^T & 0 &-T_\ell-L_\ell+M_\ell-\Lambda_\ell
\end{bmatrix} \Psi_{n\ell}=E_n^\ell \Psi_{n\ell},
\label{bdg-vortexfree}
\end{eqnarray}
with the matrix elements given by
\begin{eqnarray}
(T_\ell)_{ij}&=&-\biggl[\frac{1}{2m^*}\biggl({\beta_{i\ell}^2\over R^2}\biggr)
+\varepsilon_f\biggr] \delta_{ij}
\label{kinetic} \\
\bigl[(\Delta_\ell)_{ij},(M_\ell)_{ij}, (L_\ell)_{ij}, (\Lambda_\ell)_{ij}\bigr] &=&\int_{0}^{R}rdr\bigl[\Delta(r),{1\over2}m(r),\ell m(r),\ell\lambda_{\rm so}(r)\bigr] \phi_{i\ell}(r)\phi_{j\ell}(r).
\label{elements}
\end{eqnarray}
From the obtained spinors $\Psi_{nl}^T=(u_{1\uparrow},...,u_{N\uparrow},u_{1\downarrow},...,
u_{N\downarrow},v_{1\downarrow},...,v_{N\downarrow},-v_{1\uparrow},...,
-v_{N\uparrow})$, where the indices $n$ and $\ell$ are omitted for simplicity,
we can construct the wavefunctions $\Phi_{n\ell}(\mathbf{r})$ and solve the gap function self-consistently. We typically work with $N=200$, which is sufficient for obtaining consistent numerical results.
\begin{figure}
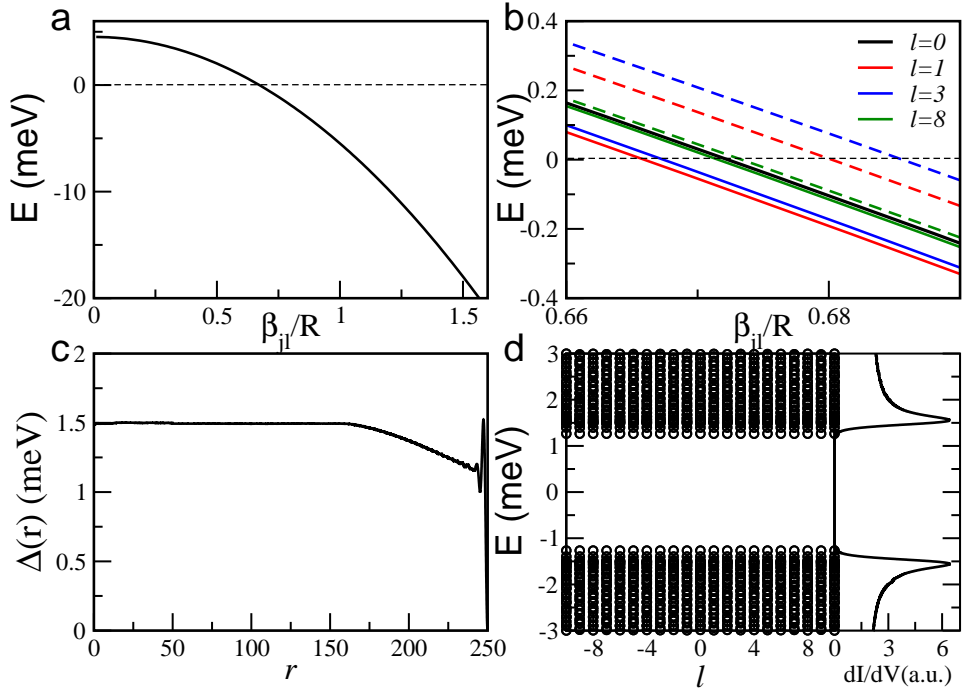

	\begin{center}
		\fig{5in}{disp.eps}\caption{(a)Free-particle dispersion $E(\beta_{j\ell})$ as function of $\beta_{j\ell}/R$. (b) Splitting of partial wave dispersions (zoomed in near Fermi level) by SOC $\lambda_{\rm so}(r)=\lambda_0\exp{(-r/r_0)}$ with $\lambda_0=20$meV and $r_0=2$. Solid (dashed) lines are for $\ell$-th partial wave carrying spin up (down). (c-d): Vortex-free SC state without magnetic ion. (c) Self-consistent pairing potential profile $\Delta(r)$. (d) Quasiparticle energy spectrum and tunneling density of states.}
	\end{center}
	\vskip-0.5cm
\end{figure}

It is instructive to note that the matrix element $T_\ell$ in Eq.~(\ref{kinetic}) describes the free-particle dispersion appearing on the diagonals of the BdG equation (\ref{bdg-vortexfree}) in the partial wave representation. Thus, the kinetic energy $E(k)=-\frac{1}{2m^*}k^2-\varepsilon_f$ has been transformed into
$E(\beta_{j\ell})=-{1\over 2m^*}({\beta_{j\ell}^2\over R^2})-\varepsilon_f$ on the disc with $\beta_{j\ell}/R$ playing the role of $k$. In Fig.S1a, the doubly spin-degenerate dispersion $E(\beta_{j\ell})$ is plotted versus $\beta_{j\ell}/R$, tracing out the hole-like band around $\Gamma$ point as partial waves of different angular momentum $\ell$ populate different points in the curve. The effects of the SOC $\lambda_{\rm so}(r)$, which gives rise to $\Lambda_\ell$ in the BdG equation, can be understood as follows. The spin and angular momentum of the partial waves are locked to form $j_z=\ell\pm{1\over2}$ states and split off from the dispersion of the unaffected $\ell=0$ channel. If $\lambda(r)$ had no spatial dependence, i.e. being a constant, the dispersions become an infinite set of equally spaced spin-split Landau levels, very much like applying opposite magnetic fields to different spin channels on a disc. However, the rapid decay of $\lambda_{\rm so}(r)$ away from the magnetic ion implies that the effects of the SOC will be limited to small but nonzero $\ell$-channels with large probability densities within the decay length $r_0$. In Fig.S1b, the calculated dispersions for $\lambda_0=20$meV are shown for the partial waves in different $\ell$ channels, zooming in close to the Fermi level. While the $\ell=1$ and $\ell=3$ states spin-orbit split away, the dispersions of the $\ell=8$ states collapse back onto the unaffected $\ell=0$ channel.

Numerically, the self-consistency process is time-consuming and limits the largest number of angular momentum channels (cutoff $L_c$) to be included. In the absence of the magnetic ion, the self-consistently determined $\Delta(r)$ using $L_c=150$ is shown in Fig.~S1c with the corresponding quasiparticle energy spectrum in Fig.~S1d for the vortex-free SC state. The paring potential profile $\Delta(r)$  begins to reduce from the uniform $\Delta$ for $r>170$, which is a consequence of the finite cutoff $L_c$ that amplifies the large-$r$ boundary effects under the disc geometry. As a result, states appear with energies just inside the expected gap energies of $\pm1.5$meV due to the ``soft boundary'' effects. This can also be seen in the rounding of the gap edge and the coherence peaks in the tunneling density of states shown in Fig.~S1d at the center of the disk. With increasing $L_c$, the boundary will become sharper and be pushed closer to the physical boundary at $R=250$. Since the physics we are interested in concerns the local effects of the SOC and exchange coupling that are limited to the small region around the magnetic ion in both the vortex and vortex-free solutions, we use the following algorithm in our numerical calculations. We first perform a self-consistent calculation in the absence of the magnetic ion using a large angular momentum cutoff $L_c=150$. The pairing function $\Delta(r)$ at large distances with $r\ge50$ is then fixed and a smaller cutoff $L_c=50$ is used in the self-consistent calculations in the presence of the magnetic ion with matching $\Delta(r)$ profile for $r\ge50$. We verified that such an algorithm improves the efficiency of the numerical computations greatly and, at the same time, ensures that our results are not affected by the boundary effects at large distances away from the center of the disk. The vortex-free solution in the presence of the magnetic ion is presented in Figs ~2g and 2h in the main text with the in-gap YSR bound states in the presence of the SOC.

\paragraph*{Vortex solutions}
Consider a single $\nu=-1$ vortex with $\mathbf{\Delta(r)}=\Delta(r)e^{-i\theta}$. The wavefunction in Eq.~(\ref{phi-mu}) becomes
\begin{eqnarray}
\Phi_{n\mu}(\mathbf{r})&=&e^{i\mu\theta}[u_{n\mu-{1\over2}\uparrow}(r)
e^{-i\frac{\theta}{2}},u_{n\mu-{1\over2}\downarrow}(r)e^{-i\frac{\theta}{2}},
v_{n\mu+{1\over2}\downarrow}(r)
e^{i\frac{\theta}{2}},-v_{n\mu+{1\over2}\uparrow}(r)e^{i\frac{\theta}{2}}]^T. \label{phi-nonrelativity}
\end{eqnarray}
Similar to the vortex-free case discussed before, the radial wave functions can be expanded using Bessel functions
\begin{eqnarray}
u_{n\mu-{1\over2}\sigma}(r)
&=&\sum_{j=1}^N u_{nj\mu-{1\over2}\sigma}
\phi_{j\mu-{1\over2}}(r), \qquad
v_{n\mu+{1\over2}\sigma}(r)=\sum_{j=1}^N
v_{nj\mu+{1\over2}\sigma}\phi_{j\mu+{1\over2}}(r),
\nonumber
\end{eqnarray}
where $\phi_{j\mu\pm{1\over2}}$ is the $j$-th order Bessel function in Eq.~(\ref{bessel}) with effective angular momenta $\mu_\pm=\mu\pm{1\over2}$ under the same cutoff $j=1,\dots, N$.
The BdG equation amounts to a $4N\times4N$ matrix eigenvalue problem
\begin{eqnarray}
\begin{bmatrix}
(T-L-M-\Lambda)_{\mu_-} & 0& \Delta_{\mu_-\mu_+} &0 \\
0 & (T-L+M+\Lambda)_{\mu_-} & 0 & \Delta_{\mu_-\mu_+} \\
\Delta_{\mu_-\mu_+}^T &0 & -(T+L+M-\Lambda)_{\mu_+}
& 0  \\
0 & \Delta_{\mu_-\mu_+}^T & 0 &-(T+L-M+\Lambda)_{\mu_+}
\end{bmatrix} \Psi_{n\mu}=E_n^{\mu} \Psi_{n\mu}
\nonumber
\end{eqnarray}
where $\Psi_{n\mu}^T=(u_{1\uparrow},...,u_{N\uparrow},u_{1\downarrow},...,
u_{N\downarrow},v_{1\downarrow},...,v_{N\downarrow},-v_{1\uparrow},...,
-v_{N\uparrow})$ with the indices $n$ and $\mu_\pm$ omitted for simplicity. Note that the BdG equation in the case of a vortex state involves both the $\mu_-$ and $\mu_+$ channels that are coupled by the pairing matrix element,
\begin{eqnarray} (\Delta_{\mu_-\mu_+})_{ij}&=&\int_{0}^{R}rdr\Delta(r)\phi_{i\mu_-}(r)
\phi_{j\mu_+}(r).
\end{eqnarray}
The rest of the matrix elements in the vortex BdG equation, i.e. $T_{\mu_\pm}$, $M_{\mu_\pm}$, $L_{\mu_\pm}$, and $\Lambda_{\mu_\pm}$, have the same expressions as the vortex-free case given in Eqs (\ref{kinetic}) and (\ref{elements}). The self-consistency procedure is the same as in the vortex-free case. The vortex solutions in both the absence and presence of the SOC and exchange coupling induced by the magnetic ion are presented in Fig.~2 and discussed in detail together with the mid-gap CdGM states in the main text.

\section{C. Vortex Solution for the electron band}

Similar to the hole band around the $\Gamma$ point, the electron band around the $M$ point in FeSe and Fe(Te,Se) superconductors can be described approximately by an electron-like parabolic dispersion $E_e(k)={1\over2m_e^*}k^2-\varepsilon_f^e$ as shown in Fig.~S2(a) by the blue-solid line. By fitting the ARPES data \cite{Skunjiang,Spzhang,Spzhang-new}, the effective mass of electron band $m_e^*\simeq1.33m_e$ with $m_e$ the free electron mass, and the Fermi energy $\epsilon_f^e\simeq25$meV. The paring gap of the electron band is $\Delta_e\simeq4$meV \cite{Smiaohu}, which can be imposed self-consistently using $g_e=64$meV and $\omega_D^e=6$meV in the BCS gap equation. Note that the impurity induced SOC is inversely proportional to the effective mass, i.e. $\lambda_{\rm so}(r)\propto {1\over m^*}$, in both sign and magnitude \cite{Selliot,Syafet}. Thus, when writing the SOC $\lambda_{\rm so}(r)=\lambda_0^e e^{-r/r_0}$ for the electron band, $\lambda_0^e$ should have an opposite sign and be scaled by the ratio of the effective mass in comparison to that for the hole band. Thus, we use $\lambda_0^e=-21$meV, while keeping $r_0$ unchanged.
\begin{figure*}
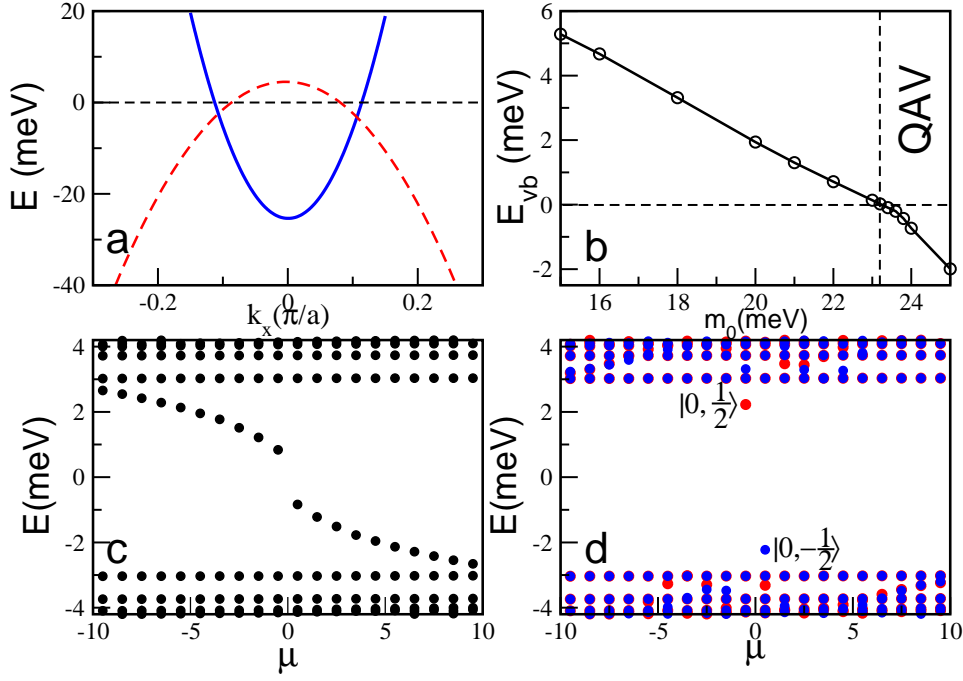

	\begin{center}
		\fig{5.0in}{electron.eps}\caption{(a) Effective bulk state electron band (blue line) near the $M$ point extracted from the ARPES data \cite{Spzhang} . The red dashed line is the effective hole band shifted from the $\Gamma$ point for comparison. (b) Electron band $\nu=1$ vortex binding energy as a function of the exchange field $m_0$ at $\lambda_0^e=-21$meV, showing the emergence of the QAV state beyond $m_0^c\simeq23.2$meV. (c) Low-energy CdGM vortex core states in the normal field-induced vortex in the absence of magnetic ion. (d) Low-energy vortex core states of the QAV induced by magnetic ion for $m_0=25$meV.}
	\end{center}
	\vskip-0.5cm
\end{figure*}

The solution of the vortex-free states for the electron band can be obtained using the same procedure discussed above for the hole band, with the corresponding substitutions of $m_e^*$, $\varepsilon_f^e$, and $\lambda_0^e$. For the vortex solution, as discussed in the main text, we need to preserve the chirality of the CdGM core states by considering a vortex in the electron band pairing order parameter $\mathbf{\Delta(r)}=\Delta(r)e^{i\nu\theta}$ with the vorticity $\nu=1$, which is opposite to that of the hole band. The wavefunction in Eq.~(\ref{phi-mu}) becomes
\begin{eqnarray}
\Phi_{n\mu}(\mathbf{r})&=&e^{i\mu\theta}[u_{n\mu+{1\over2}\uparrow}(r)
e^{i\frac{\theta}{2}},u_{n\mu+{1\over2}\downarrow}(r)e^{i\frac{\theta}{2}},
v_{n\mu-{1\over2}\downarrow}(r)
e^{-i\frac{\theta}{2}},-v_{n\mu-{1\over2}\uparrow}(r)e^{-i\frac{\theta}{2}}]^T. \label{phi-nonrelativity}
\end{eqnarray}
Similar to the case for the hole band, the radial wave functions can be expanded in the basis of Bessel functions.
The BdG equation amounts to a $4N\times4N$ matrix eigenvalue problem
\begin{eqnarray}
\begin{bmatrix}
(T-L-M-\Lambda)_{\mu_+} & 0& \Delta_{\mu_+\mu_-} &0 \\
0 & (T-L+M+\Lambda)_{\mu_+} & 0 & \Delta_{\mu_+\mu_-} \\
\Delta_{\mu_+\mu_-}^T &0 & -(T+L+M-\Lambda)_{\mu_-}
& 0  \\
0 & \Delta_{\mu_+\mu_-}^T & 0 &-(T+L-M+\Lambda)_{\mu_-}
\end{bmatrix} \Psi_{n\mu}=E_n^{\mu} \Psi_{n\mu}
\nonumber
\end{eqnarray}
where $\Psi_{n\mu}^T=(u_{1\uparrow},...,u_{N\uparrow},u_{1\downarrow},...,
u_{N\downarrow},v_{1\downarrow},...,v_{N\downarrow},-v_{1\uparrow},...,
-v_{N\uparrow})$ with the indices $n$ and $\mu_\pm$ omitted for simplicity. The BdG equation in the presence of a vortex involves both the $\mu_-$ and $\mu_+$ channels that are coupled by the pairing matrix element,
\begin{eqnarray} (\Delta_{\mu_+\mu_-})_{ij}&=&\int_{0}^{R}rdr\Delta(r)\phi_{i\mu_+}(r)
\phi_{j\mu_-}(r).
\end{eqnarray}
The matrix elements for the kinetic energy is given by
\begin{equation} (T_{\mu_\pm})_{ij}=\biggl[\frac{1}{2m_e^*}\biggl({\beta_{i\mu_\pm}^2\over R^2}\biggr)
-\varepsilon_f^e\biggr] \delta_{ij}.
\end{equation}
The rest of the matrix elements in the vortex BdG equation, i.e. $M_{\mu_\pm}$, $L_{\mu_\pm}$, and $\Lambda_{\mu_\pm}$, have the same expressions as given in Eq.~(\ref{elements}).
In Fig.~S2b, the calculated vortex binding energy is shown as a function of the exchange field $m_0$ for the electron band.
The QAV from the electron band emerges beyond $m_0^{c,e}\simeq23.2$meV. Fig.~S2c shows the vortex core CdGM states in the normal field induced vortex in the absence of the magnetic ion. These low-energy vortex core states are pushed into the continuum by the exchange field in the QAV nucleated at the magnetic ion, as shown in Fig.~S2d for $m_0=25$meV.

\section{D. Helical Dirac fermion TTS coupled to quantum anomalous
	vortices}
Finally, we discuss the electron-doped TSS coupled to the QAV. The helical Dirac fermions carry an extra Berry phase \cite{Ssharlai99,Sniu-rmp10} since $\partial_x\pm i\partial_y=\exp{(\pm i\theta)}(\partial_r\pm i\partial_\theta/r)$ in the dispersion.
The corresponding wavefunction in Eq.~(\ref{phi-mu}) for a single $\nu=1$ vortex is therefore given by
\begin{eqnarray}	\Phi_{n\mu}(\mathbf{r})=e^{i\mu\theta}[u_{n\mu\uparrow}(r),u_{n\mu+1\downarrow}(r)
e^{i\theta},v_{n\mu-1\downarrow}(r)e^{-i\theta},-v_{n\mu\uparrow}(r)]^T. \label{phi-relativity}
\end{eqnarray}
Due to the combination of the vorticity-induced phase and the Berry phase, the quantization condition reflected in Eq.~(\ref{phi-relativity}) now requires the quantum number $\mu$ to be an integer, i.e. $\mu=0,\pm1,\pm2,\dots$. This is the crucial difference compared to the vortex wavefunction for the parabolic bulk band discussed before where $\mu$ is a half-integer. The integer $\mu$ allows the presence of zero-energy mode in the CdGM states.
The BdG equation (\ref{bdg}) contains the kinetic part $H_{\rm kin}^\prime$ given in Eq.~(\ref{hkin-tss}), $H_{\rm soc}^\prime=\lambda_{\rm so}^\prime(r)L_z\sigma_z$ and $H_{\rm ex}^\prime=-m^\prime(r)(L_z+{\sigma_z\over2})$. Moreover, the BdG equation for the Dirac fermions will mix $\mu$ and $\mu\pm1$ channels of the wavefunction in Eq.~(\ref{phi-relativity}). Expanding the radial parts using the Bessel functions as before,
\begin{eqnarray}
u_{n(\mu,\mu+1)\sigma}(r)&=&\sum_{j=1}^Nu_{nj(\mu,\mu+1)\sigma}
\phi_{j(\mu,\mu+1)}(r),  \\
\qquad v_{n(\mu,\mu-1)\sigma}(r)&=&\sum_{j=1}^Nv_{nj(\mu,\mu-1)\sigma}
\phi_{j(\mu,\mu-1)}(r),
\end{eqnarray}
we obtain the BdG equation as a $4N\times4N$ matrix eigenvalue problem
\begin{eqnarray}
\begin{bmatrix}
-(L+M-\Lambda)_\mu-\varepsilon_f'& V_{\mu,\mu+1} & \Delta_{\mu,\mu-1} &0 \\
V_{\mu,\mu+1}^T& -(L-M+\Lambda)_{\mu+1}-\varepsilon_f' & 0 & \Delta_{\mu+1,\mu} \\
\Delta_{\mu,\mu-1}^T &0 & -(L+M+\Lambda)_{\mu-1}+\varepsilon_f'& -V_{\mu-1,\mu}  \\
0 & \Delta_{\mu+1,\mu}^T & -V_{\mu-1,\mu}^T & -(L-M-\Lambda)_\mu+\varepsilon_f'
\end{bmatrix} \Psi_{n\mu}=E_n^\mu \Psi_{n\mu}.
\nonumber
\end{eqnarray}
The matrix elements in the above BdG equation are given by
\begin{eqnarray}
&&(V_{\mu,\mu^\prime})_{ij}=\frac{2v_D}{R}
\frac{\beta_{i\mu}\beta_{j\mu^\prime}}{\beta_{i\mu}^2-\beta_{j\mu^\prime}^2}, \qquad
(\Delta_{\mu,\mu^\prime})_{ij}=\int_{0}^{R}rdr\Delta_{\rm QAV}(r)
\phi_{i\mu}(r)\phi_{j\mu^\prime}(r),
\label{dirac-kinetic} \\
&&\bigl[(M_\mu)_{ij}, (L_\mu)_{ij}, (\Lambda_\mu)_{ij}\bigr] =\int_{0}^{R}rdr\bigl[{1\over2}m^\prime(r),\mu m^\prime (r),\mu\lambda_{\rm so}^\prime(r)\bigr] \phi_{i\mu}(r)\phi_{j\mu}(r),
\label{dirac-elements}
\end{eqnarray}
and $\Psi_{n\mu}^T=(u_{1\uparrow},...,u_{N\uparrow},
u_{1\downarrow},..,u_{N\downarrow},v_{1\downarrow},..,v_{N\downarrow},
-v_{1\uparrow},...,-v_{N\uparrow})$  with the indices $n,\mu,\mu\pm1$ omitted for simplicity. In Eq.~(\ref{dirac-kinetic}), $\Delta_{\rm QAV}(r)$ is the pairing profile of the QAV generated by the magnetic ion from the bulk states.

The energy spectrum of the TSS and the LDOS at the magnetic ion are shown in Fig.~3 and compared to the STM tunneling conductance measured at the excess ion sites in Fe(Te,Se) in the main text. The localized mode at zero energy corresponds to $\mu=0$. It is associated with the creation operator $\gamma_0^\dagger$. From Eq.~(\ref{gamma})
\begin{eqnarray}
\gamma_{0}^\dagger&=&\int d^2\mathbf{r} \bigl [ u_{0\uparrow}(r)\psi_{\uparrow}^{\dagger}(\mathbf{r})
+v_{0\uparrow}(r)\psi_{\uparrow}(\mathbf{r}) +u_{0\downarrow}(r)e^{i\theta}
\psi_{\downarrow}^{\dagger}
(\mathbf{r})+v_{0\downarrow}(r)e^{-i\theta}
\psi_{\downarrow}(\mathbf{r})\bigr],
\end{eqnarray}
where the localized wavefunctions of the zero energy mode $u_{0\uparrow}(r)$, $v_{0\uparrow}(r)$, $u_{0\downarrow}(r)$, $v_{0\downarrow}(r)$ are plotted in Fig.S3 a-d. Their behaviors clearly show that $u_{0\uparrow}(r)=v_{0\uparrow}(r)$ and $u_{0\downarrow}(r)=v_{0\downarrow}(r)$. As a consequence,
\begin{equation}
\gamma_{0}^\dagger=\gamma_{0},
\end{equation}
indicating that the localized zero energy mode is a charge neutral Majorana zero-mode.
\begin{figure}
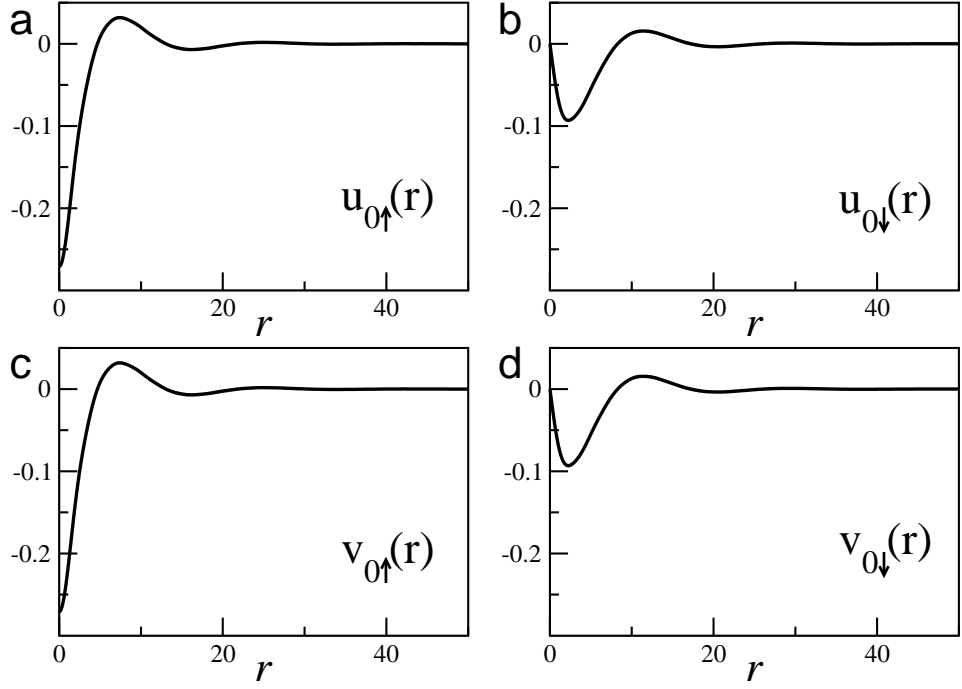

	\begin{center}
		\fig{5in}{uv.eps}\caption{The localized wavefunctions of the zero energy mode (a) $u_{0\uparrow}(r)$, (b) $u_{0\downarrow}(r)$, (c) $v_{0\uparrow}(r)$, (d) $v_{0\downarrow}(r)$.}
	\end{center}
	\vskip-0.5cm
\end{figure}

The nonmagnetic part of the impurity potential $U(r)$ has not been included in the present study since the effects of a time-reversal invariant potential in $s$-wave superconductors are known to be weak and do not lead to qualitatively new physics. We verified that including $U(r)$ only makes the mid-gap states more localized spatially, but does not change qualitatively the obtained results.

\end{widetext}

\end{document}